\documentclass{article}
\usepackage{here,graphicx,subfigure,latexsym}

\begin{document}
\title{\boldmath Development of the MRPC for the TOF system of the MultiPurpose Detector}


\author{
V.A.~Babkin,
S.N.~Bazylev,
I.S.~Burdenyuk,
M.G.~Buryakov,\\
A.V.~Dmitriev,
P.O.~Dulov,
V.M.~Golovatyuk,
S.P.~Lobastov,\\
M.M.~Rumyantsev,
A.V.~Schipunov,
A.V.~Shutov,
I.V.~Slepnev,\\
V.M.~Slepnev,
A.V.~Terletskiy,
S.V.~Volgin\\
\em{Joint Institute for Nuclear Research, Joliot-Curie st. 6, Dubna, Russia}
}

\maketitle

\begin{abstract}The Multipurpose Detector (MPD) \cite{cite1.MPD_CDR} is designed to study of hot and dense baryonic matter 
in collisions of heavy ions over the atomic mass range 1~--~197 at the centre of mass energy up to $\sqrt{S_{NN}}$ = 11~GeV 
(for Au79+). The MPD experiment will be carried out at the JINR accelerator complex NICA \cite{cite2.NICA_CDR} which 
is under construction. The barrel part of the MPD consists of various detectors surrounding the interaction point.  
It includes a precise tracking system (time projection chamber (TPC) and silicon inner tracker (IT)) and high-performance 
particle identification system based on time-of-flight (TOF) and calorimeter (ECal). The triple-stack multigap resistive plate chamber 
is chosen as an active element of the TOF. It provides good time resolution and long term stability. 

This article presents parameters of the MRPC obtained using the deuteron beam of JINR accelerator Nuclotron. 
The time resolution is $\sim$40~ps with efficiency of 99\%. Rate capability studies resulted with a time resolution 
of 60 ps and efficiency higher than 90\% on the beam with particle flux densities up to 2 kHz/cm$^2$.
\end{abstract}




\flushbottom

\section{Design of the MPD time-of-flight system}
\label{sec1:TOF_design}

To meet its physics goals, the MPD requires good particle identification capabilities covering a large phase 
space. Identification of charged hadrons in the momenta range (0.1 -- 2~GeV/c) is achieved by
using time-of-flight and momentum measurements as well as the energy loss ($dE/dx$) 
from the TPC and IT detector systems.

The basic requirements for the TOF system are:

-- high granularity in order to keep the overall system occupancy below 15\%;

-- good position resolution to provide effective matching of the TOF hits with the TPC tracks;

-- high geometrical efficiency;

-- separation of pions and kaons up to $p_{t}$ < 1.5 GeV/c;

-- separation of (anti)protons up to $p_{t}$ < 3 GeV/c; and

-- TOF detector elements must function in the magnetic field up to 0.5 T.

The barrel part of the TOF is located in the MPD between the time-projection chamber 
and the electromagnetic calorimeter and has internal radius of 1.45~m from the beam axis. 
The barrel is segmented into 14 sectors with the length of 5.9~m. It covers the pseudorapidity 
range $|\eta| \le$ 1.4. One sector contains two independent modules. 
The module consists of two separate volumes: the inner gas box which contains 10 MRPCs arranged 
at the angle of 6 degrees to the beam axis; the outer box contains front-end electronics (FEE) cards, cables, high 
voltage and gas plugs. One MRPC with active surface of 640$\times$300~mm$^2$ has 24 strips of 10 mm 
width with pitch of 12.5 mm. Signals are read out from both sides of the strip. The  number of the MRPCs 
in the barrel is 280. The number of FEE channels is 13440. The total surface of the barrel part of the TOF system is $\sim$52 m$^2$. 
The occupancy does not exceed 15\% and the geometric efficiency of the TOF detectors in this arrangement is $\sim93\%$.


\begin{figure}[tp]
\centering 
\includegraphics[width=0.45\textwidth]{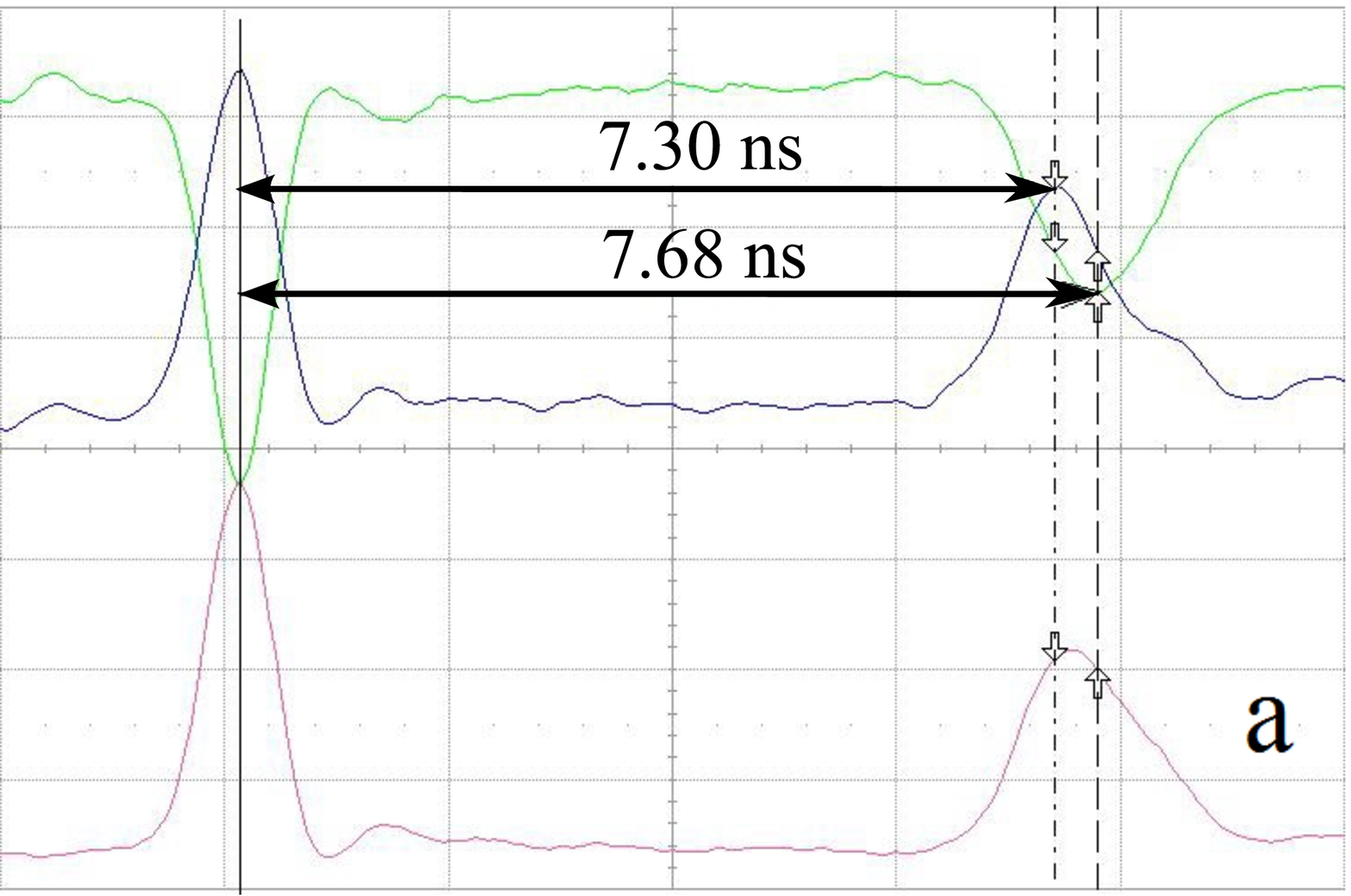}
\hfill
\includegraphics[width=0.45\textwidth]{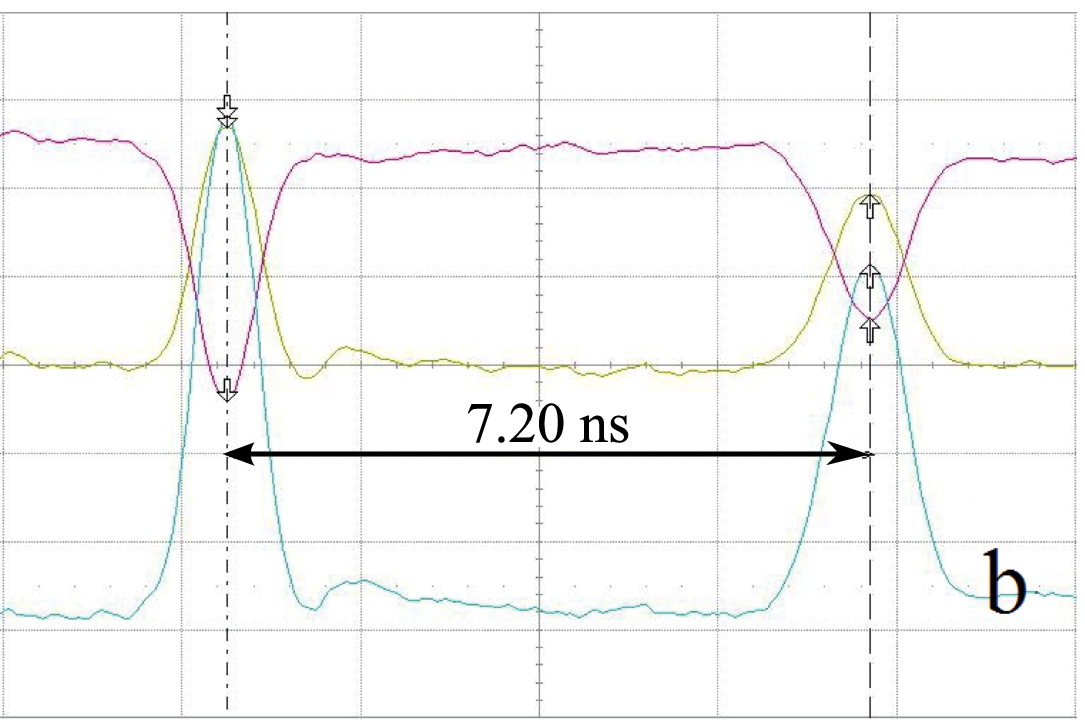}
\caption{\label{fig0:ds}Reflections of the differential signal for the:
(a)~double-stack and (b)~tripple-stack MRPC.}
\end{figure}

The MRPC provides high resolution by using the time-over-threshold method only if the analog signal from the detector is minimally 
distorted on the input of preamplifier. Therefore one needs to provide very good termination of 
impedance between the strip, the cable and the preamplifier input. Any reflection could cause the wrong estimation of the signal 
width when the amplitude correction is applied.
Working with the double stack MRPC with differential signal readout scheme one has to pay attention to the fact that internal and 
external electrodes have different impedances. The inner strip is surrounded from both sides by the fiberglass PCB and glass with 
dielectric constant $\varepsilon\sim$~4. The external electrodes have a honeycomb on one side and glass on the other. 
As a result the positive and negative parts of the signal propagate with different velocities (figure~\ref{fig0:ds}a) 
\cite{cite3.Babkin_POS}. This fact can lead to a worsening of time resolution. In order to reduce the dispersion effect of the 
positive and negative part of differential signal, the detector design must be symmetrical.

\begin{figure}[bp]
\centering 
\includegraphics[width=0.85\textwidth]{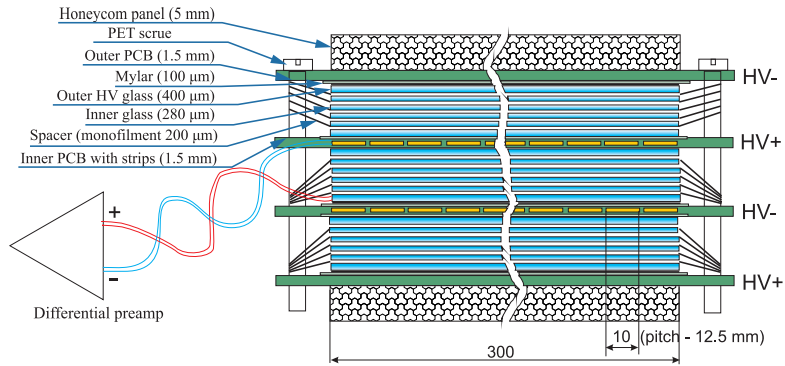}
\caption{\label{fig1:tsrpc_scheme}Scheme of the triple-stack MRPC.}
\end{figure}

In order to have a large signal and symmetrical geometry we decided to use a triple-stack design of multigap resistive plate chamber 
(figure~\ref{fig1:tsrpc_scheme})~\cite{cite4.Babkin_TS_RPC}. Each stack of the detector consists of 5 gas gaps with the width of 200~$\mu$m. 
It has glass resistive electrodes with the thickness of 280 $\mu$m. Thin glass was chosen to improve the rate capability of the detector 
\cite{cite5.Weiping}. High voltage is applied to the external glass of each stack. For this purpose, the outer 
surface of the glass is coated with a conductive coating with a surface resistance of 5 -- 20 M$\Omega/\Box$. 
Readout strips in the detector are located only on the inner PCBs. Outer PCBs do not have any conductors. 
The copper layer inside the cover of the gas box used as a ground in this case. It was made from the following considerations.

If we include strips on the outer PCBs and connect them with the inner strips, the impedance of the differential transmission line of 
three stacks becomes very low ($\sim18~\Omega$).The minimum input impedance of the NINO ASIC \cite{cite6.Usenko} is 30~$\Omega$.
In order to match the impedances of the strip and the NINO input is necessary to increase impedance of the transmission line twice, for example by 
adding a serial resistor of $18~\Omega$. In turn, the current at the amplifier input will be reduced by half.
Another problem of such readout configuration is that the signal velocity propagation on the 
outer strip is higher than those on the inner strip and as consequence the dispersion of differential signal.
When we use outer strips as a ground, the impedance of differential line corresponds to the impedance of the twisted pair. 
On the other hand, crosstalk significant increases due to strong capacitive coupling.

Terefore we decided to use outer PCB boards without conductor, and to have readout strips only on the inner PCBs. One pair of the inner strips 
(anode and cathode) has the differential impedance of 55 $\Omega$ and well match with parallel connected twin twisted pair cable 
with the same impedance without termination resistors. This design has no problem with the reflection of signals from the edges of the strip. 
The cathode and anode strips are in identical environment of conductors and dielectrics. Speeds of propagation of the 
signal along the cathode and anode are equal in such a transmission line (figure~\ref{fig0:ds}b). Thus, the dispersion of the signal is avoided. 
Because high voltage is applied to all three stacks, signals from the avalanches induce on the inner strips both from the internal 
and external stacks. Other advantage of this design is that the large number of narrow gaps improve the time resolution of the 
detector \cite{cite7.24gapMRPC}. 


The disadvantage of the triple-stack MRPC is that we read only the signal of one polarity from the outer stack. Therefore we lose about one 
third of the total charge produced in all gaps of the triple-stack MRPC.

\section{Readout electronics for the TOF system}
\label{sec2:Electronics}

\begin{figure}[bp]
\centering 
\begin{minipage}[h]{0.44\textwidth}
\includegraphics[width=1\textwidth]{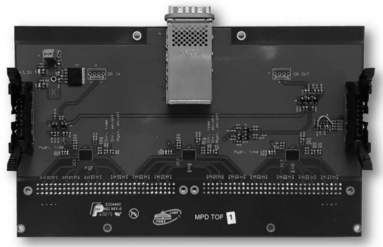}
\caption{\label{fig2:NINO}NINO based 24-channel amplifier-discriminator with the Molex CXP  connector.}
\end{minipage}
\hfill
\begin{minipage}[h]{0.55\textwidth}
\includegraphics[width=1\textwidth]{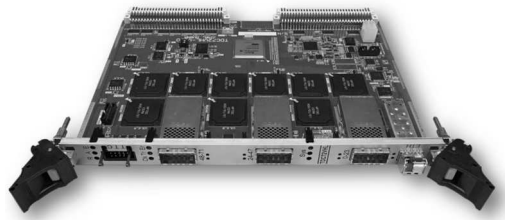}
\caption{\label{fig3:TDC72VHL}72-channel time-to-digital converter TDC72VHL based on HPTDC.}
\end{minipage}
\end{figure}

The 24-channel front-end board (figure~\ref{fig2:NINO}) based on the NINO ASIC was designed. 

It has the following features:

-- stabilization of the supplied voltage;

-- overload protection of input channels;

-- capacitors on the inputs for the double-end strip readout; and

-- monitoring and control of the sensitivity threshold by the slow control system.

The MPD TOF preamplifier-discriminator is adapted for the long strip readout. To prevent reflection of the signal 
at the inputs, the preamplifier provides the ability to fine-tune the input impedance for matching of the differential transmission 
line of the chamber and preamplifier.

The overall dimensions of the preamplifier board are 200~$\times$~120~mm$^2$. The voltage stabilizer is installed on the 
board and eliminates the voltage drop on the cable. The preamplifier can be powered with a voltage of 3 to 6~V. The output signal is in 
the LVDS standard. The preamplifier has the Molex CXP output connector constructed by InfiniBand network technology. 
The measured time resolution of one channel of the preamplifier is $\sim$7~ps. 

A new VME64x time-to-digital converter TDC72VHL (figure\ref{fig3:TDC72VHL}) based on the HPTDC chip was designed for the MPD TOF 
readout. One VME module has 72 differential 100~$\Omega$ LVDS inputs. Three amplifier boards can be connected to 
one such module. The time-sampling of the TDC72VHL is 24.4 ps per bin. The TDC72VHL provides the ability of the precise 
White Rabbit~\cite{cite8.WR} synchronization with other timing devices and can operate in a standalone mode.

The HPTDC chip has a high integral nonlinearity (INL). We used the code density test to calibrate the VME TDC72VHL modules. 
The time resolution of one channel of readout electronics (including the NINO time jitter) is better than 20 ps after the INL calibration.

\section{Beam test results}
\label{sec3:Test_res}

\begin{figure}[bp]
\centering 
\includegraphics[width=1.0\textwidth]{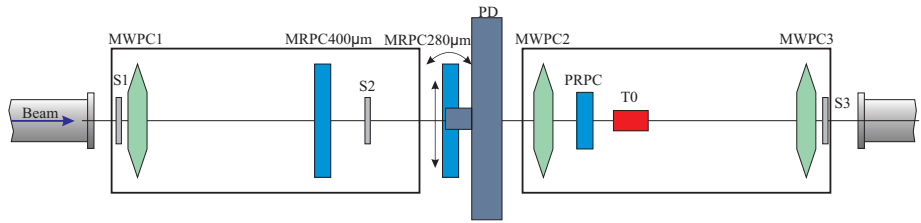}
\caption{\label{fig4:Test_MPD}Scheme of the test setup at the ``Test beam MPD''.}
\end{figure}

The triple-stack MRPC prototype was tested using the 3.5 GeV/A deuteron beam from the Nuclotron at the ``MPD test beam'' 
facility at JINR (figure~\ref{fig4:Test_MPD}). 
The test setup consisted of:

-- three proportional chambers (MWPC1--3) each with 6 coordinate planes for tracking and definition of the beam profile with an accuracy of 1 mm; 

-- three scintillation counters (S1--S3) for trigger and control of the beam intensity; 

-- fast start detector (T0) based on the MCP-PMT XP82015/A1Q from Photonis with time resolution of 40 ps\cite{cite9.Yurevich};

-- the precision positioning device (PD) for movement and rotation of the tested detector related to the beam axis operated by remote control; and

-- data acquisition system (DAQ) based on the standard VME and Ethernet.

\begin{figure}[tp]
\centering 
\includegraphics[width=0.51\textwidth]{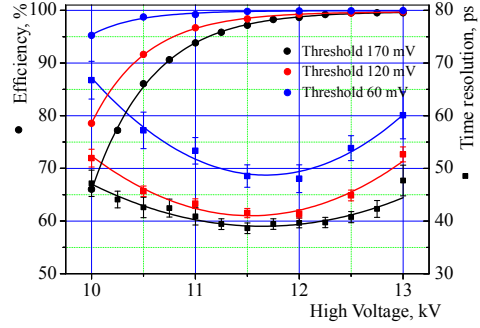}
\caption{\label{fig5:eff_tr_280}Efficiency and time resolution of the MRPC prototype for different NINO thresholds.}
\end{figure}

The dependence of time resolution and efficiency on the applied high voltage for different NINO discriminator thresholds are presented 
in figure~\ref{fig5:eff_tr_280}. The best time resolution for the tested prototype is 40 ps with efficiency higher than 98\%. 

\begin{figure}[bp]
\centering 
\begin{minipage}[h]{0.48\textwidth}
\includegraphics[width=1.\textwidth]{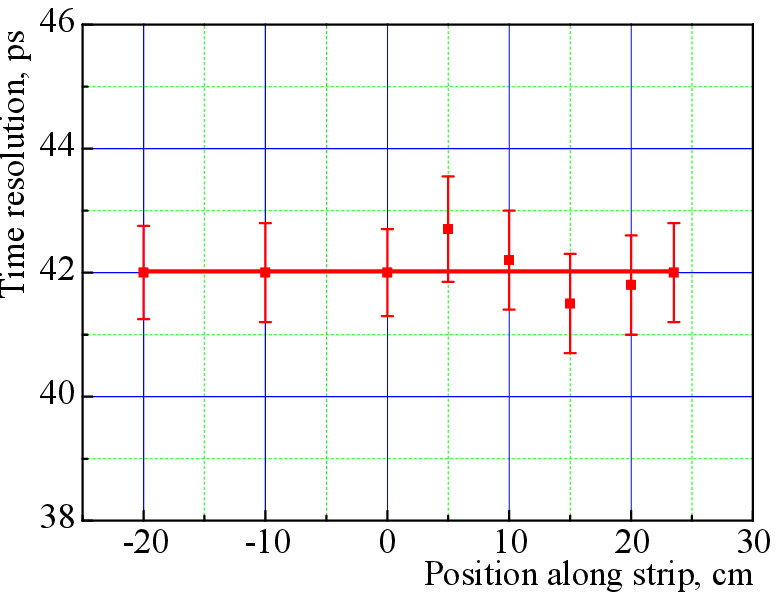}
\caption{\label{fig6:tr_along}Dependence of time resolution on the position of particles along the strip.}
\end{minipage}
\hfill
\begin{minipage}[h]{0.49\textwidth}
\includegraphics[width=1.\textwidth]{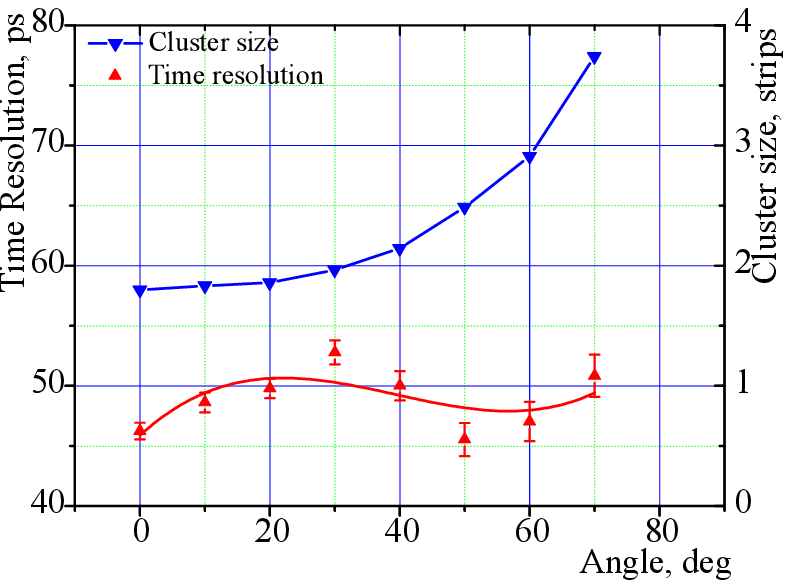}
\caption{\label{fig7:tr_angle}Time resolution and cluster size dependence on the angle of particles.}
\end{minipage}
\end{figure}

The positioning device allows a movement of the detector in the $X$ and $Y$ directions with an accuracy of 20 $\mu$m and 
to rotate in two planes ($XZ$ and $YZ$). 
The time resolution along the strip was measured by moving the MRPC in a horizontal direction. (figure~\ref{fig6:tr_along}).

A GEANT simulation shows that secondary particles could enter the TOF detectors of the MPD at angles up to 60 degrees. 
Dependence of time resolution on the angle between the beam direction and the line accross the strip is presented in figure~\ref{fig7:tr_angle}. 
During turn of the detector some slight changes were observed, but time resolution did not exceed 50 ps.

\begin{figure}[tp]
\centering 
\begin{minipage}[h]{0.49\textwidth}
\includegraphics[width=0.99\textwidth]{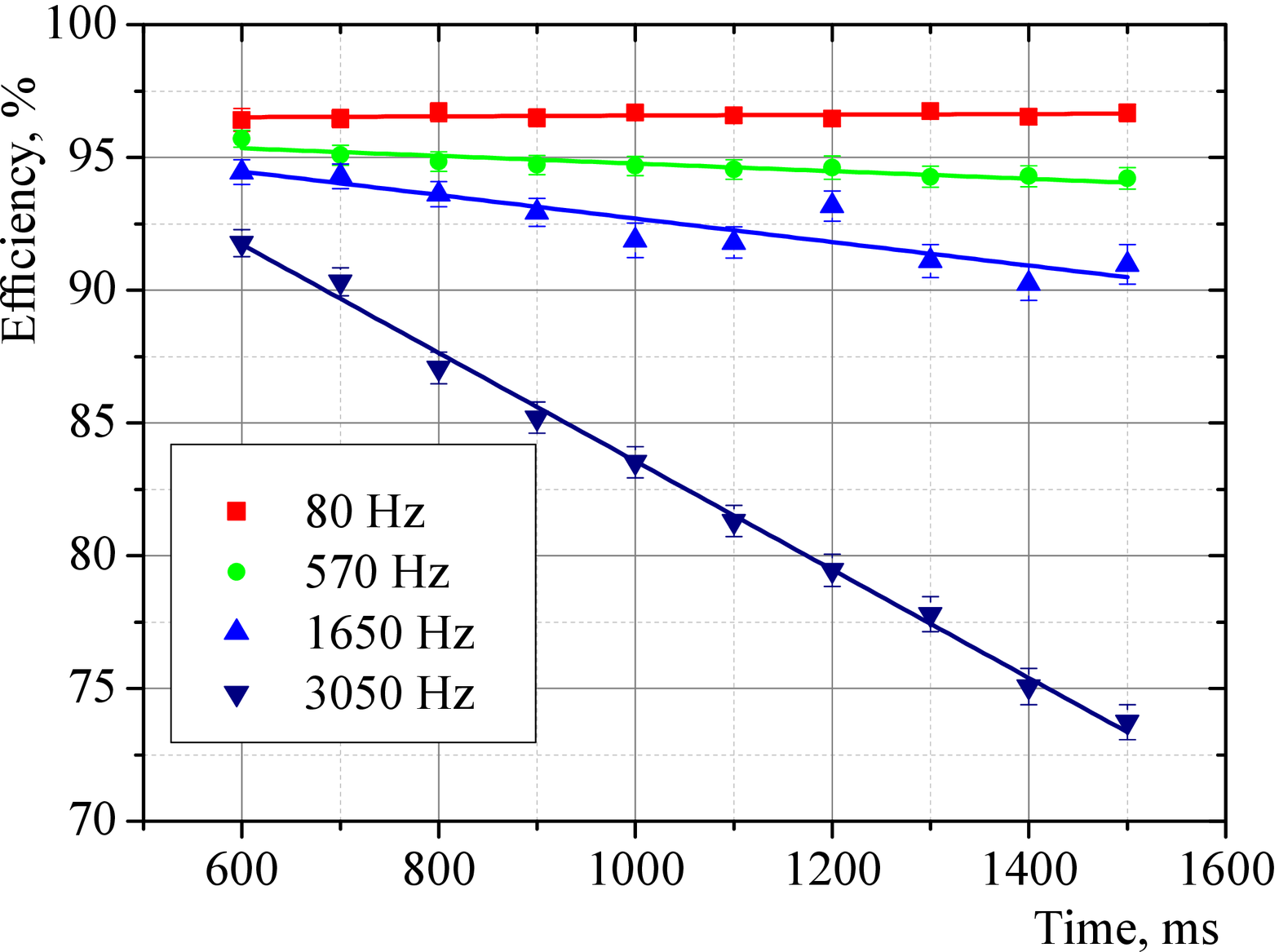}
\caption{\label{fig8:eff_intime} Time degradation of the detector efficiency due to particles flux.}
\end{minipage}
\hfill
\begin{minipage}[h]{0.49\textwidth}
\includegraphics[width=0.99\textwidth]{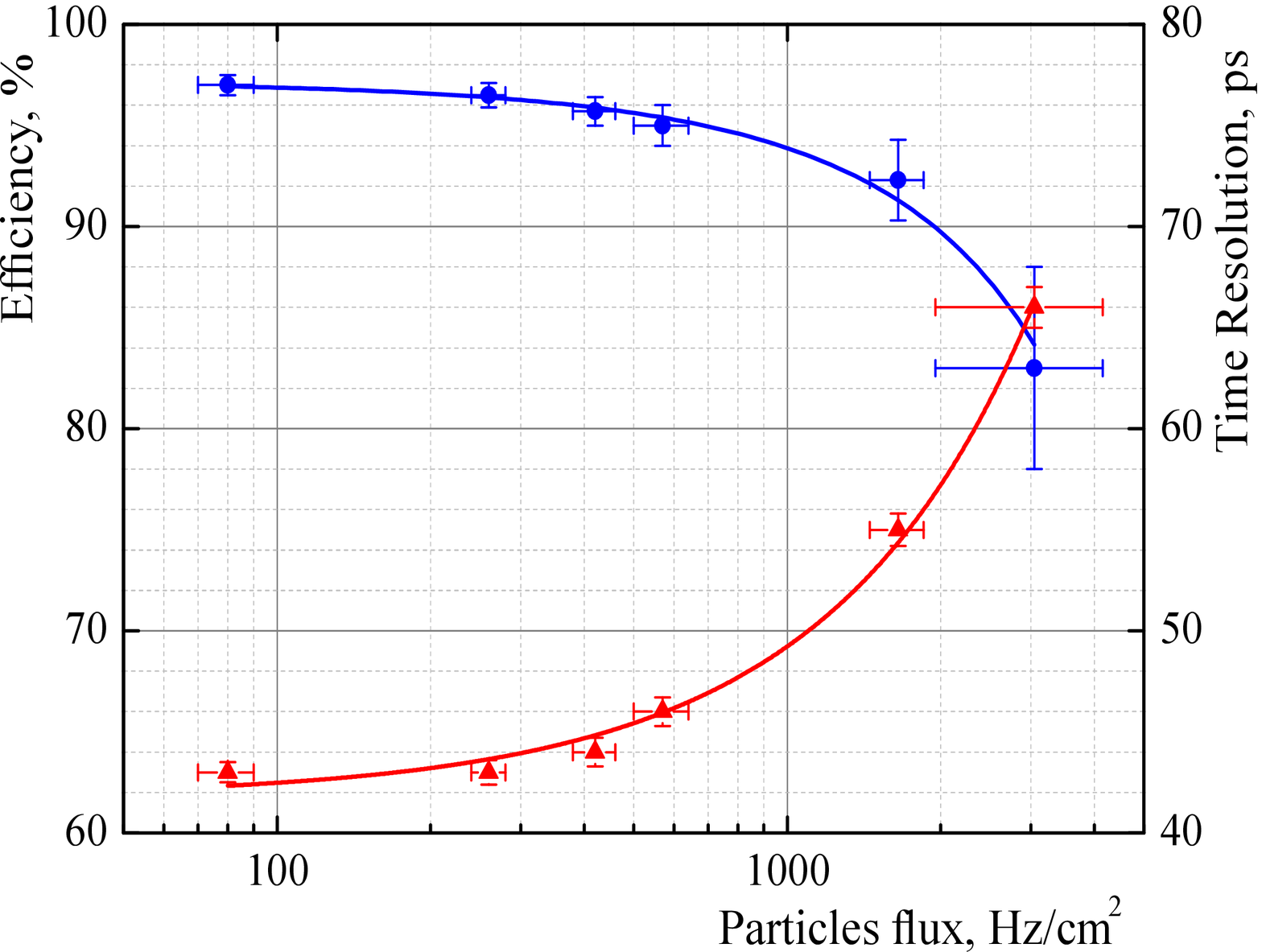}
\caption{\label{fig9:ave_rate}Average efficiency and time resolution depending on intensity.}
\end{minipage}
\end{figure}

We have also studied the rate capability of the triple-stack MRPC.
Particle rate was determined by the scintillation counter which was read out to the multi hit scaler MSC16V with time sampling of 1 $\mu$s. 
We have observed that the efficiency and time resolution decreased during irradiation (figure~\ref{fig8:eff_intime}). 
The efficiency dropped from 92\% to 74\% during 1 second when the flux of deuterons is about 3~kHz/cm$^2$. 
The plot of the MRPC efficiency and time resolution as a function of average over 1 second particle flux is presented in figure~\ref{fig9:ave_rate}.

\section{Conclusions}
\label{sec4:Conclusions}

The triple-stack MRPC with thin (280 $\mu$m) glass was chosen as the main element of the TOF system. Time resolution of the system which 
consists of MRPC and T0 ($\sigma_{t} \sim$ 40 ps) is better than 60 ps and satisfies the MPD requirements. 
It has been shown that the time resolution of the MRPC is almost independent of the position along the strip and the angle of the particle trajectory
to the strip.

\end{document}